\documentclass[aps,prb,twocolumn,groupedaddress]{revtex4}
\usepackage{graphicx}
\usepackage{amssymb}
\usepackage{bm}
\usepackage{subfigure}
\usepackage{graphpap}
\begin{document}

\title{ Electron channels in biomolecular nanowires}
\author{Arrigo Calzolari*,$^1$
Rosa Di Felice,$^1$ Elisa Molinari,$^1$ and
Anna Garbesi$^2$ }

\affiliation{ $^1$ INFM-S$^3$ - National Research Center on
nanoStructures and Biosystems at Surfaces, and Dipartimento di Fisica
Universit\`a di Modena e Reggio Emilia, I-41100 Modena, Italy\\
$^2$ CNR ISOF, Area della Ricerca, Bologna, Italy}


\begin{abstract}
We report a first-principle study of the electronic and conduction properties 
of a quadruple-helix guanine wire (G4-wire), a DNA-derivative, with inner 
potassium ions. The analysis of the electronic structure highlights the 
presence of energy manifolds that are equivalent to the bands of 
(semi)conducting materials, and reveals the formation of extended electron 
channels available for charge transport along the wire. 
The specific metal-nucleobase interactions affect the electronic properties 
at the Fermi level, leading the wire to behave as an intrinsically p-doped system.
\end{abstract}

\maketitle
*email: calzolari.arrigo@unimore.it

\section{Introduction} 

Fueled by the ever increasing drive for miniaturization and improved performance in 
electronic devices and by potential applications in the nanotechnologies, the research 
effort to investigate the properties of novel nanowire materials is undergoing an 
impressive growth. It is foreseen that the conventional solid-state technology could 
be replaced by new generations of devices based on molecular components, which take 
advantage of the quantum mechanical effects that rule the nanometer scale. 
Molecular electronics~\cite{1} is currently explored as a long-term alternative for 
increasing the device density in integrated circuits. However, to keep up with these 
expectations, the new trend must provide flexible, reproducible and well structured 
architectures, easy to wire in a programmable way.  

By virtue of their recognition and self-assembling properties, DNA molecules seem 
particularly suitable to fulfill these requirements. Both the intrinsic combinatory 
principles of nucleic acids and their chemistry can be exploited to build precise, 
miniaturized and locally modulated patterns, where the drawing of functional arrays 
is obtained through a series of programmed chemical reactions and not by the physical 
handling of the samples. The realization of DNA-based wired architectures via self-assembly 
is a viable route to scale down the size of devices to the molecular level.~\cite{2,3,4} 
However, whereas it was demonstrated that the self-assembling capabilities of DNA make it 
suitable as a template to wire metallic materials,~\cite{5} its ability as an intrinsic conductor 
is questioned by experimental results.~\cite{6} Depending on base sequence, molecule length, 
environmental conditions, substrates, and electrode materials, the direct measurements of 
the dc conductivity of DNA-based structures in solid-state devices~\cite{7} report insulating 
character,~\cite{8} semiconductor-like transport characteristics,~\cite{9} ohmic behavior,~\cite{10} and 
proximity-induced superconductivity.~\cite{11} 
Indeed, even in the cases in which charge transport 
has been observed, the current is very low, with resistances of  the order of 0.1-1 G$\Omega$ 
cross the length of the DNA molecules (variable between 10 $\mu$m and 10 nm). 
Thus, due to its apparent poor intrinsic conductivity, DNA might be reasonably considered 
as a bad insulator rather than a viable electrical molecular wire.

Besides the standard DNA, other nucleotide-based helical molecules, such as guanine 
quadruple helices (G4-wires) or "metal-manipulated" duplexes, may offer the desired 
mechanical, recognition, and self-assembly properties, that make DNA so attractive. 
With respect to native DNA,  these derivatives have metal cations in the inner core of the base stack. 
Whereas the interactions between {\em external} ions and the double helix have been largely studied both 
experimentally and theoretically, the effects of their inclusion inside the helix are largely unknown. 
The presence of {\em internal} metal ions may drastically affect the bonding pattern with and among the bases, 
introducing novel features in the structural and electronic properties of the system.~\cite{12} 
A promising pathway for the exploitation of DNA as a conductor in molecular devices is indicated 
by the evidence that metal ions incorporated in the helix core may modify the conductivity of DNA-based wires.~\cite{13,14}
\noindent
\begin{figure}[!t]
\includegraphics[clip,width=0.50\textwidth]{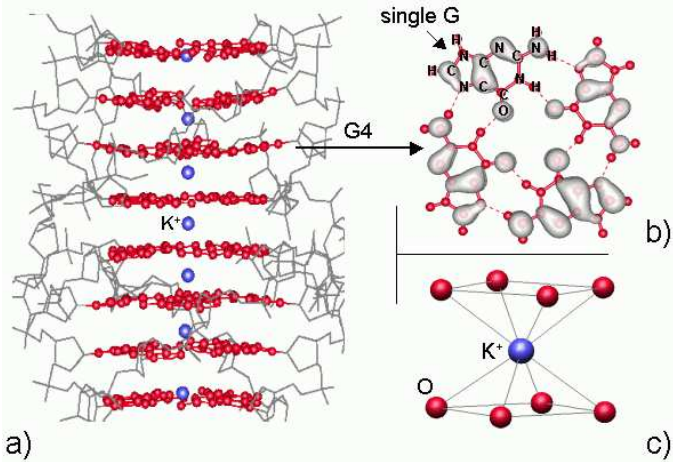}
\caption{
(a) Side view of a short quadruplex from which the simulated G4-wire was extracted. 
The image was derived from an experimental X-ray structure.~\cite{21} 
The central core of the helix (red) is constituted of stacked supramolecular structures, the G-quartets (G4), and the K$^+$
ions (blue spheres) are hosted in the central cavity of the G-quartet stack. 
The lateral gray sticks describe the external sugar-phosphate backbone, excluded in our simulations. By virtue of the overall C$_4$ 
symmetry of the helix, it is possible to extract a periodic unit of three G4 
tetrads from this finite quadruplex, that constitutes the elementary building-block of the simulated infinite wires. 
(b) Isolated G-quartet that forms one plane of the stacked G4-wire. Each G4 consists of four co-planar 
H-bonded guanines, arranged in a square-like configuration. The image shows the relaxed geometry and an 
isosurface plot of the HOMO manifold. (c) Scheme of the bypiramidal K$^+$-O coordination.
}
\end{figure}

In this paper, we focus on nanowires known as G4-wires~\cite{15} (or {\em quadruplexes}), consisting of stacked 
guanine (G) tetrads (G4): the structure of these systems is illustrated and described in Figure 1. 
These one-dimensional polymers are becoming appealing as prospective candidates for bio-molecular 
electronics because, due to the low ionization potential of guanine (the lowest among nucleic-acid bases), 
they might be suitable to mediate charge transport by hole conduction along the helix, and have even 
been suggested as nano-mechanical extension-contraction machines.~\cite{16} 
In the presence of appropriate metal cations (especially K$^+$ and Na$^+$), solutions of homoguanylic 
strands in water,~\cite{17,18} as well as lipophilic guanosine monomers in organic solvents,~\cite{19} self-assemble 
in right-handed quadruple helices. The G4-motif has been identified in both cases; however, while the 
quadruplexes obtained from guanylate strands have an outer mantle of sugars and phosphates (as in DNA) 
that connects the adjacent G4 planes, those obtained from lipophilic guanosine derivatives  have 
no interplanar connection: a continuous linker between consecutive tetramers is not necessary to form the wires. 
The guanine quadruplexes have recently attracted interest because of their possible role in biological systems:~\cite{20} 
their biological relevance has propelled a large number of investigations (e.g. X-ray~\cite{21} and NMR~\cite{22}) aimed 
at characterizing such unusual supramolecular structures. Quadruple helices have been obtained in the presence 
of different monovalent (K$^+$, Na$^+$, NH4$^+$)~\cite{19}$^{a,}$~\cite{23} and divalent 
(Ca$^{2+}$, Ba$^{2+}$, Sr$^{2+}$, Pb$^{2+}$)~\cite{19}$^{b,}$~\cite{24} cations. 
Despite the different chemical nature of the constituents, all these G4-wires are characterized by an 
inner core of stacked G4's intercalated by metal cations: one cation in {\em every} tetrad and one cation in 
{\em every other} tetrad, in the case of mono- and di-valent ions, respectively. 
The self-assembly capability of the G4-helices allows for the formation of quite long (100-1000 nm) and stiff wires. 
AFM images~\cite{25} of four-stranded helices obtained from G-rich oligonucleotides on substrates revealed 
the G4-wires to be uniform and relatively rigid polymers, with few bends, kinks, or branches. 

Despite the large amount of  structural investigations, the conduction properties of these nanowires 
are  basically unknown and a direct measurement of electrical properties of G4-wires is still missing. 
From the theoretical point of view, quantum chemistry and molecular dynamics studies focused on the 
energetics and on the geometry of isolated G-quartets or finite clusters of stacked G4's,~\cite{26} 
whereas the electronic properties of these materials were so far investigated to a much lesser extent.~\cite{27} 

In the following, we present a first-principle investigation of the electronic and conduction properties 
of periodically repeated G4-wires. Recently,~\cite{27} we described the structure and the energetics, as well as 
some basic features of the electronic structure, of an infinite G4-wire with and without the presence of K$^+$ 
ions in the inner cavity. In this paper, we focus entirely on the electronic properties of the same system 
and present a complete and thorough analysis of the guanine-guanine and metal-guanine interactions, 
discussing the effects on the conduction properties of the tubular system.

\section{Computational approach}

We performed ab-initio calculations of the electronic and conduction properties of infinite G4-wires 
in the presence of K$^+$ ions, within the Density Functional Theory (DFT) approach,~\cite{28} using the PW91~\cite{29} 
gradient-corrected exchange-correlation functional. DFT is lately gaining a large credit in the 
scientific community as a reliable and accurate method to describe large-scale biomolecular 
aggregates,~\cite{30} including guanine-based stacks.~\cite{31} 

Our total-energy-and-force calculations~\cite{32} allowed us to attain a simultaneous description of the 
optimized atomic configuration and of the corresponding electronic structure for the selected 
systems (see Figure 1): the finite planar guanine tetrad (G-quartet) and the infinite helical 
G4-wire filled with K$^+$ ions in the inner cavity (label 3G4/K$^+$). The single-particle electron 
wave-functions were expanded in a plane-wave basis set with a kinetic-energy cutoff of 25 Ry. 
Two special {\bf k}-points in the irreducible wedge were employed for Brillouin Zone (BZ) sums in 
the case of the G4-wire. The infinite helix was simulated by repeated supercell containing three 
stacked G4 tetrads,~\cite{27} employing periodic boundary conditions in the three spatial direction 
a thick vacuum layer ($\sim 16$ \AA) in the directions perpendicular to the helical axis 
prevented spurious interactions between adjacent replicas of the wire. For the isolated G-quartet, 
the same vacuum thickness was employed also in the third direction perperdicular to the plane 
of the tetrad, and only the $\Gamma$ point was used in the BZ sampling. 

The electron-ion interaction was described by non-norm-conserving pseudopotentials~\cite{33} for all 
the species (C, N, O, H) except K, for which a norm-conserving pseudopotential~\cite{34} was used. 
For the latter species, both the valence $4s$ and the semi-core $3p$ shells contributed to the 
system with valence electrons. This treatment represented a significant refinement towards a 
complete description of the electronic structure, with respect to the simplified results~\cite{27} 
obtained by fixing the $3p$ shell in the frozen core and by applying {\em Non Linear Core Corrections}~\cite{35} 
(NLCC) to account for partial core relaxation.

The starting atomic configuration was obtained from the results of the X-ray analysis~\cite{21} of 
the d(TG4T) quadruple helix. Motivated by the observation that G4-wires form with~\cite{17,18} and without~\cite{19} 
the covalent skeleton and by our specific interest in the base stack as a channel for charge mobility, 
we neglected the external backbone in our simulations and focused on the central core of the helix, 
constituted of guanines and metal cations (see Fig. 1a). This choice is supported by theoretical 
reports~\cite{36} for approximated DNA structural models, which assert that if any current flows in such 
systems, it does through the base-stacking, without involving the external mantle in the transport phenomena. 
The same evidence was more recently confirmed by DFT calculations of the electronic properties of 
real DNA sequences (A-DNA~\cite{8}$^a$ and Z-DNA~\cite{37}): these simulations showed that the orbitals related to 
the sugar-phosphate backbone are a few eV below the Highest Occupied Molecular Orbital (HOMO) and 
above the Lowest Unoccupied Molecular Orbital (LUMO) of the system.  
In principle, one should expect that the environment surrounding the base stack plays an important 
role in the overall conductivity properties of DNA molecules.~\cite{37,38,39} Recent studies~\cite{37,39}$^b$ pointed 
out that a random distribution of counter-ions in the unit cell modifies the electrical properties 
of  DNA both reducing the {\em bulk} quantum conductance and introducing localized empty states in the energy gap.
However, this result is coherent with a picture where the external ions {\em quantitatively} affect the 
global transport properties of the system, but do not constitute an alternative pathway for the 
electron/hole transport through the helix.~\cite{7} Therefore, we believe that the {\em bare} guanine core will 
be well representative of the essential electronic properties of the quadruplexes, allowing for the 
inspection of the charge-migration mechanisms and of the key features of metal-molecule interaction. 
An explicit account for the effects of the backbone and of the surrounding counterions~\cite{37,39} would 
instead be demanded for a quantitative evaluation of the quantum conductance to be compared with 
measured transport characteristics, which is way beyond the purpose of the present work.

\section{Results and discussions}

By means of the first-principle approach outlined above, we optimized both the isolated 
G-quartet and the periodic G4-stack. 

The analysis of the electronic structure of the planar G-quartet shows 
interesting features that will help understanding the electrical properties of G4-wires. 
As also found for the other planar G-aggregates (e.g. dimers, ribbons),~\cite{31} the H-bonds among the 
guanines do not favor the formation of supramolecular orbitals extended on the whole G-quartet 
and the existence of dispersive bands. On the other hand, the intermolecular interactions split 
each guanine energy level into a multiplet structure: each multiplet is composed of four 
(the number of G's in the tetrad) energy levels and has a total width of about 200 meV. 
The orbitals that contribute to a manifold have identical character and are localized on the
individual G molecules. Figure 1b shows an isosurface plot of the convolution of four electron 
states (the HOMO's of the four guanines in the tetrad) which form the $\pi$-like HOMO manifold. 

By exploiting the square-symmetry of the planar tetrad and the 30$^{\circ}$ 
twist angle between consecutive tetrads, we simulated a quadruplex of infinite length 
with a periodically repeated unit supercell containing three stacked G-quartets and three 
intercalated K$^+$ ions in the unit cell.~\cite{27}
Each potassium ion was symmetrically located between 
two consecutive tetrads, and bipyramidally coordinated with the eight (four above and four below) 
nearest-neighboring oxygen atoms (Fig. 1c). The atomic positions were relaxed until the forces 
vanished, within an accuracy of 0.03 eV/\AA.
The structure, the energetics, and the metal-induced stability of the tube were described elsewhere.~\cite{27} 
We find now that the explicit inclusion of the semi-core $3p$ electrons of K in the valence shell for 
the pseudopotential calculations does not alter the results obtained previously within the NLCC approximation,~\cite{27} 
and does not change the understanding of the system from the structural point of view. 
The refined treatment of the semi-core $3p$ electrons of K allows us to gain a deeper insight into the electronic structure.
\noindent
\begin{figure}[!t]
\includegraphics[clip,width=0.50\textwidth]{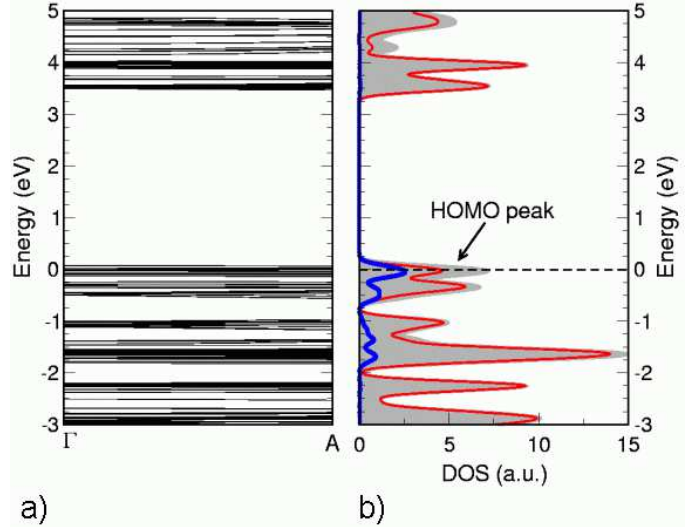}
\caption{
(a) Band-structure along the GA direction parallel to the axis of the 3G4/K$^+$ G4-wire, 
which coincides with the direction along which the planar tetrads stack. (b) Total and Projected Density of States (DOS) 
of the wire. The shaded gray area represents the total DOS. The thin red (thick blue) solid curve marks 
the projection of the DOS onto G (K). Note that the total DOS and the G-DOS coincide in a large part 
of the spectrum, where the thin red curve almost coincides with the border of the shaded area. 
The dashed horizontal curve indicates the Fermi level.
}
\end{figure}

The band-structure of the G4-wire 3G4/K$^+$ is shown in Figure 2a: the unoccupied states (above 3.5 eV) 
are separated from the occupied states (below 0 eV) by a large energy gap (affected by the typical DFT 
underestimation~\cite{28}), and some of the states around zero energy are partially filled. 
We previously demonstrated~\cite{31} that $\pi$-$\pi$
coupling among guanines may give rise to delocalized Bloch-type orbitals, whose band dispersion 
along the stacking direction depends on the relative rotation angle between nucleobases in adjacent planes. 
Taking into account the detected rotation angle~\cite{21} of 30$^{\circ}$
for the guanine quadruple helices, we now report that the electronic band-structure of the 
3G4/K$^+$ column is represented by dispersionless energy bands, indicating that no complete 
supramolecular orbital delocalization is to be expected for such a degree of helicity. 
This behavior leads to the conclusion that in G4-wires the $\pi$-$\pi$ 
superposition is not sufficient to induce a sizable coupling between the molecular states 
of guanines and/or the atomic orbitals of potassium, which would lead to {\em purely} dispersive bands 
similar to those of (semi)conducting materials. However, as for the isolated G-quartet, the 
band-structure of the 3G4/K$^+$ column identifies the presence of manifolds that are equivalent to 
effective bands: each manifold contains a number (or multiplet) of levels that can be explained in terms 
of the number of nucleobase molecules and of potassium ions in the periodic supercell, as we discuss later. 
The orbitals in a multiplet have identical symmetries and some degree of hybridization between neighboring bases. 
The levels within a manifold are so close (dense) in energy ($\Delta$E $\le$ 20meV) that an external weak 
interaction (e.g. the thermal fluctuations or an  exterinal electric field) may let the orbitals mix, 
giving an {\em effective} dispersive band. This interpretation suggests a description of the electronic structure 
where the spreading of the energy levels of the manifold leads to the formation of {\em effective} band-like peaks 
in the Density of States (DOS) (Fig. 2b, shaded gray area). 
The DOS of the 3G4/K$^+$ wire thus appears similar to those of materials with a dispersive band-structure. 

It is important to note that there is a one-to-one correspondence between the band structure and the 
ground state conduction properties (e.g., the quantum conductance spectrum) of a periodic system in 
the coherent transport regime:~\cite{40} at any given energy, the quantum conductance is a constant value 
proportional to the number of transmitting channels available for charge mobility, which are equal 
(in the absence of external leads) to the number of bands at the same energy. 
Therefore, the formation of effective dispersive bands in the investigated G4-wire is a signal of 
the {\em intrinsic} capability of the material of hosting electron "energy channels" available for charge 
migration, whithin continuous energy ranges. 
On the contrary, the energy spikes that would characterize a discrete spectrum would not be efficient 
for transport. The extended orbitals (see below) related to the effective bands turn out to be the 
corresponding viable "space channels" for carrier mobility, e.g. the pathways through which the 
carriers may migrate in the wire. 

A comparison between the manifolds of the isolated G-quartet and of the G4-wire highlights an 
increased density of energy levels in the manifolds of the quadruple helix with respect to the tetrad. 
This effect is due to two factors: (i) the number of energy levels in each multiplet; (ii) the in-plane 
and out-of-plane guanine-guanine and guanine-metal interactions, that couple the electron states stemming 
from the various structural elements (either from G or from K) in the supercell. 
The number of electron states in each manifold depends upon the number of molecules and ions in the unit cell: 
therefore, whereas each manifold of the G-quartet contains four levels, the manifolds of the 3G4/K$^+$ wire 
contain twelve levels from the twelve guanines, and additional twelve levels from the K$^+$ ions in the energy 
range where metal-base hybridization occurs. We come back to this point later when we discuss the DOS.
The number of levels in a manifold is not the only element that determines the details of the bandstructure: 
another key feature is the band-width, which is instead controlled by the specific interactions. 
Whereas in the case of the G4 tetrad only H-bonding plays a role, in the case of the 3G4/K$^+$ wire 
there are different contributions from H-bonding and stacking between the bases, and from the coupling between 
the bases and the metal ions. It is worth noting that the formation of dense manifolds and the consequent 
establishment of {\em effective band-like} potential conduction channels is a common characteristic of stacked 
H-bonded nucleobase aggregates. In fact, not only we identified the same features in guanine quadruplexes 
both in the presence and absence of inner metal cations,~\cite{27} but similar energy manifolds were also detected  
in the electronic band-structure of two different DNA duplexes.~\cite{7,8}$^{a,}$~\cite{37} 
For instance, in their simulation of 
an eleven base-pair poly(dG)-poly(dC)~\cite{8}$^a$ sequence, de Pablo and coworkers found a HOMO-manifold derived  from 
the eleven states of guanines. In that case, the topmost valence band had a bandwidth of only 40 meV, smaller 
than that calculated in our G4-wire ($\sim$ 700 meV). 
This confirms that, while the splitting of the energy levels into multiplets is a fingerprint 
of stacked rotated H-bonded nucleobases, the peculiar characteristic of  the manifolds depend 
on the intrinsic stacking properties (e.g. the $\pi$-$\pi$ coupling) of each type of helix.

To investigate more closely the effects of the metal cations and of the metal-molecule interaction, 
we projected~\cite{41} the total density of states of 3G4/K$^+$ (shaded gray area in Fig. 2b) on G (thin red curve) 
and K (thick blue curve) atomic orbitals. The G- and K-Projected DOS's (PDOS's) show that in a large 
portion of the spectrum (below -2 eV and above 3.5 eV) only the guanine electron states contribute 
to the DOS. For instance, the LUMO peak at 3.5 eV has a {\em pure} guanine character, being the convolution 
of the only twelve LUMO orbitals of twelve G's. The contribution of the K$^+$ ions to the total DOS is 
mainly localized at the top of the valence band (between -2 eV and 0 eV). 
The presence of both guanine and potassium contributions to the DOS (as seen from the superposition 
of the blue and red peaks in Fig. 2b) at the top of the occupied bands is the consequence of a 
metal-molecule interaction. The HOMO peak of the 3G4/K$^+$ wire (see the double-peak indicated by an 
arrow in Figure 2b) is due to a convolution of 24 electron states deriving from the coupling 
between the twelve G HOMO's and the twelve potassium orbitals. 
Indeed, one would expect that the K$^+$ ions would contribute to the DOS only with the filled $p$ 
orbitals, after the $s$ electrons are removed upon ionization of the system.
However, if this were the case, by counting the levels in the manifolds we would find only nine 
completely occupied levels due to the metal (three $p$ orbitals from each K ion), in addition to 
the guanine-based manifolds containing twelve levels each. Instead, the number of computed occupied 
levels (twelve more than those coming from the guanine counting, and all concentrated in the HOMO 
double-peak) and the occurrence of partial occupation for some of them (those around the computed 
Fermi level, see Figure 2b) indicates $sp$ hybridization in potassium when inserted in the helical guanine complex. 
Therefore, potassium does not contribute separately with the $4s$ and the filled $3p$ shells, but with a 
partially occupied $sp$ shell: 7 equivalent electrons in 4 orbitals, which become 6 electrons in 4 
orbitals when the system is ionized as the 3G4/K$^+$ wire. 
Consequently, the highest-energy component of the HOMO double-peak has an occupation factor of 3/4
the Fermi level is pinned at the top of the effective valence band (Fig. 2b) and  the G4-wire in the 
presence of K$^+$ ions behaves as an intrinsically p-type doped system. 
The observation of delocalized holes at the topmost valence band is in agreement with the description 
of the electronic structure of potassium discussed above, where each ionized K$^+$ atom contributes to 
the whole electronic structure not with three fully occupied $p$-levels, but with four partially occupied $sp$-states.
\begin{figure}[!t]
\includegraphics[clip,width=0.50\textwidth]{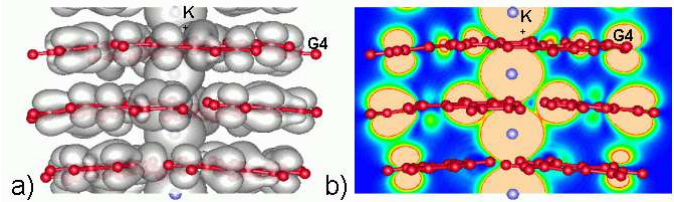}
\caption{
(a) Isosurface plot of the convolution of orbitals that represents the HOMO manifold of the G4-wire, side view. 
(b) Contour plot of the same orbital in a plane perpendicular to the tetrameric planes and containing the K$^+$ ions.
}
\end{figure}

The spread $sp$-orbitals of potassium easily interact with the surrounding molecular orbitals of the tetrads, 
giving {\em hybrid} metal-molecule states. We visualized this effect by drawing the convolution of the effective 
HOMO of the wire. Figure 3a shows an isosurface of the HOMO charge density in the unit cell, obtained from 
the convolution of the 24 highest occupied orbitals that contribute to the HOMO double-peak. 
By comparison with Figure 1b, it is possible to recognize a uniform distribution of the $p$ orbitals deriving 
from the tetrads, but also the inner delocalized density stemming from the orbitals of the K$^+$ ions. 
By cutting the HOMO charge density with a plane perpendicular to the G-quartets, we further inspect 
the features of the metal-molecule interaction (Fig. 3b). 
Due to the high degree of charge delocalization along the wire axis, the G4-wires may be described 
as good electron/hole channels for mobile charges. Two types of pathways for such mobile charges 
can be identified in Figure 3b and contribute to the conductivity channels. 
The first one stems from the low-energy component of the HOMO double-peak, is extended through the 
guanine core of the helix, and is due to the base-base interaction. 
Similar channels are also observed for the other manifold-derived peaks (e.g. LUMO) 
and in the empty (K free) quadruple helix.~\cite{27} 
The second type of pathway is due to the high-energy component of the HOMO peak and results from 
the metal-base interaction. This hybrid pathway is centered around the potassium ions in the central 
cavity of the wire, and it clearly shows the coupling with the coordinated oxygen atoms of the G 
molecules. Hence, the inclusion of the metal cations inside the helix drastically influences the 
electronic properties of the system. The details of the metal-nucleobase interactions depend upon 
the nature of the cation included in the helix. In the present case of potassium, its coupling with 
the G bases modifies the topmost valence band, favoring the formation of extended orbitals along the 
stacking direction and enhancing the conduction properties of the empty guanine structure.~\cite{27} 

To get further insights into the metal-base coupling, we compared the 3G4/K$^+$ wire with other two 
similar neutral systems: the empty quadruple helix (labeled 3G4) and the G4-wire with neutral K 
atoms (labeled 3G4/K). The structures 3G4 and 3G4/K do not describe real systems, but are useful 
models for better understanding the 3G4/K$^+$ quadruplexes. The changes of the inner core do not 
basically modify the geometry (bond lengths and angles) of the quadruple helix, but affect the 
electronic structure. Figure 4 shows the comparison among the DOS (shaded gray areas) and the 
PDOS (thin red and thick blue curves) of the systems. 
To compare the three structures we aligned the bottom energy levels, and fixed the origin of 
the energy scale at the top of the valence band of the empty 3G4 helix (taken as reference), 
with the purpose to outline the effects of the inclusion of metals into the empty guanine 
supramolecular structure.

As already mentioned, in the DOS of all the studied G4-wires we observe the occurrence of 
broad G-derived peaks, reflecting the same guanine aggregation state. The empty tube has the 
valence band completely occupied and the Fermi level lays in the middle of the gap; 
in the presence of  potassium (3G4/K and 3G4/K$^+$), the Fermi level is pinned at the top of a 
partially occupied valence band. In the 3G4 case, the HOMO peak  is the convolution of  the 
HOMO's of guanines (${\bf H_G}$ in Fig. 4), which are completely occupied; in the other two cases, instead, 
the HOMO derives from the interactions between the HOMO's of guanine (${\bf H_G}$) and the hybrid states 
of potassium (labeled ${\bf K}$ in Fig. 4), which are partially filled. One may consider the 3G4/K 
structure as obtained by adding three electrons to the 3G4/K$^+$. 
The inclusion of such electrons increases (from 3/4 to 7/8 filling factor) but does not 
complete the occupation of the topmost band; thus, the helix with the atomic K 
would also be an intrinsic p-type doped wire.
\begin{figure}[!t]
\includegraphics[clip,width=0.45\textwidth]{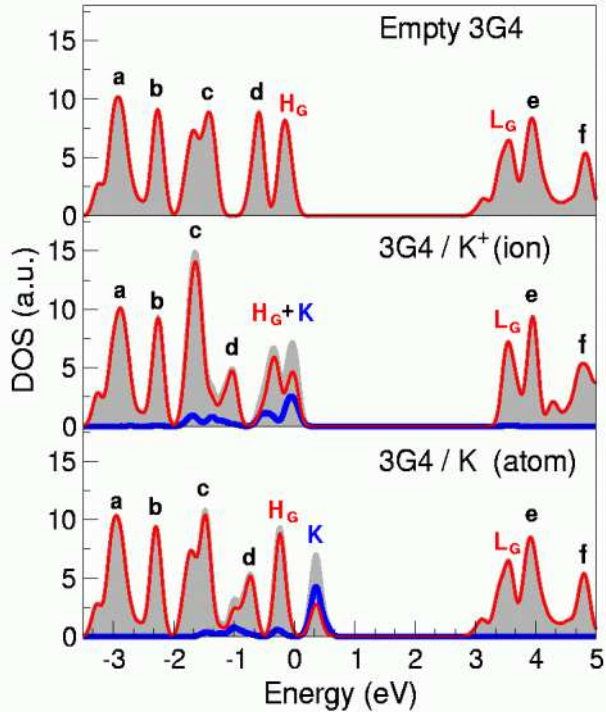}
\caption{
Total and Projected Density of States (DOS) of different quadruple helical G4-wires: 
without inner metal (3G4) in the top panel, with K$^+$ ions (3G4/K$^+$) in the middle panel, 
with neutral K atoms (3G4/K) in the bottom panel. In each panel, the shaded gray area 
represents the total DOS, the thin red (thick blue) curve marks the contribution of 
guanine (potassium) to the DOS. ${\bf H_G}$ (${\bf L_G}$) labels the band-like peak deriving from the 
HOMO's (LUMO's) of single guanine molecules in the unit cell. The label ${\bf K}$ identifies 
peaks with a pronounced potassium contribution. 
The origin of the energy scale is consistently set at the HOMO of the empty column in 
the three plots. The 3G4 is an insulator with a large HOMO-LUMO gap. 
In the metal-containing quadruplexes 3G4/K$^+$ and 3G4/K, instead, the Fermi level is 
pinned at the top of the HG+K and of the K peaks, respectively, due to partially 
occupied electron states. 
}
\end{figure}

By analyzing the position and the width of the band-like peaks for the three systems, 
we can underline further details about the K-G interactions. 
As the K-derived states are found at the top of the valence band, far away from that range 
the spectra are identical, e.g. peaks ${\bf a, b, e, f}$  have the same width, shape and 
positions in the three plots. Also the G-like LUMO peak ${\bf L_G}$ remains basically unchanged. 
On the contrary, in the 3G4/K$^+$ wire (central panel in Figure 4) the occurrence of a 
non-vanishing K-PDOS in the [-2;-1] eV range strongly modifies the peaks ${\bf c}$ and ${\bf d}$ with respect 
to the empty wire: the electrostatic coupling between the $s$-like peak ${\bf d}$ and the $sp$ orbitals of 
the K$^+$ ions shifts the position of this peak to lower energies and reduces its height. 
All the other peaks in Fig. 4 have a $p$-like character and are less influenced by the G-K$^+$ coupling. 
The displacement of peak ${\bf d}$ also modifies peak ${\bf c}$, which is sharper and higher in 3G4/K$^+$ than 
in the free-standing tube. In the neutral 3G4/K helix, the electrostatic coupling between K 
and G is more effectively screened because of the larger number of electrons in the system: 
as a consequence, the shift of peak ${\bf d}$ is reduced with respect to the 3G4/K$^+$ case, and peak ${\bf d}$ 
remains similar to that of the empty wire. Another difference that we remark is the composition 
of the HOMO peaks: whereas for the 3G4 empty tube the HOMO is purely G-like (${\bf H_G}$), it is a mixed 
G-K-like double-peak (${\bf H_G+K}$) for the charged 3G4/K$^+$ wire. For the neutral 3G4/K wire, the HOMO  
(Fig. 4, middle panel) is split into two sharper peaks (Fig. 4, bottom panel): a lower energy 
component similar to the peak  ${\bf H_G}$ and a higher component with a prevalent ${\bf K}$ character. 
The decrease in the degree of K-G orbital mixing in 3G4/K with respect to 3G4/K$^+$ is most likely 
attributable to the increased occupation of the $sp$ shell of potassium, 
which becomes more inert because more similar to a closed shell. 

These changes in the electronic structure due to different charge states of the inner metals 
underline the delicate equilibrium that rules the stability and the mutual interactions in this 
hybrid organic/inorganic molecule/metal complex. 
Differently from M-DNA,13 where the metal cation substitutes for an immino hydrogen atom 
and is covalently bonded to a nucleobase, in G4-wires there is no direct charge sharing 
between the metal and the guanines. The stability and the electronic properties of the system 
are ruled by the coordination ratio among the K$^+$ and the eight nearest-neighboring oxygen atoms, 
symmetrically located above and below the ion (Fig. 1c). 
The metal cations stabilize the structure via electrostatic interactions, but also constitute an 
{\em effective bridge} between the electronic charge density around the oxygen atoms: 
the HOMO contour plot shown in Figure 3b is the result of the $\pi$-$\pi$
coupling between the oxygens of  consecutive tetrads, mediated by the electronic structure of the K$^+$ ion. 

The possibility of changing the electronic properties of the G4-wire, by using different metal 
ions to stabilize the stack, is expected to be a powerful tool for tuning the conduction properties of 
the nanowires. We are currently exploring the inclusion of other cations with different outer-shell 
electronic configurations (e.g. the transition metals), that may be exploited to further change 
the conductivity of G4-wires. 

\section{Conclusions}

The first principle study of the electronic and conduction properties of  infinite G4-wires, stabilized 
by K$^+$ inner cations, shows that the $\pi$-$\pi$ coupling among stacked planar tetrads is insufficient 
to induce the formation of dispersive bands in the wires. However, the presence of closely spaced 
energy levels leads to the formation of manifolds, whose density of states suggests a band-like behavior. 
The coupling among guanine-localized molecular orbitals, which may be easily induced by a weak external 
interaction, gives rise to extended electron channels, suitable to host mobile charge carriers along the wire.

While the formation of split manifolds seems to be a general feature of base-base interactions in 
H-bonded stacked supramolecular nucleobase aggregates, the effects due to the presence of  the metals 
depend on the identity of the cations. In the case of potassium, the inclusion of the cations enhances 
the conduction properties of the system, generating additional extended electronic channels stemming from 
the metal-guanine interaction. The mixed $sp$-orbitals of K$^+$ hybridize with the HOMO's of guanine, and this 
coupling gives origin to a partially filled HOMO band, which makes the system equivalent to a p-doped 
wide-bandgap semiconductor. 

The above results, along with their attractive mechanical and self-assembly properties, suggest that 
G4-wires may be explored as viable DNA-based conductors for nanoscale molecular electronics. 
The computed properties are the equivalent of those of a bulk material: of course, the actual behavior as a 
wire in a device setting depend on the specific device implementation, is affected from conditions such as 
electrode-wire and substrate-wire coupling, and leaves several open issues for further investigation and exploitation.

\section{Acknowledgments}

We gratefully thank Joshua Jortner for illuminating discussions. This work was partially supported 
by the EC through project "DNA-based nanowires" IST-2001-38951, by INFM through "Progetto calcolo parallelo" 
which provided computer time at CINECA (Bologna, Italy), and by MIUR (Italy) through grant "FIRB-NOMADE".


\begin{thebibliography}
\bibitem{41}The G projection was obtained by summing over the projections onto the atomic orbitals of all 
            the species contained in the guanine molecules.
\end{thebibliography}
\end{document}